\documentclass[aps,prd,notitlepage,twocolumn,
superscriptaddress, preprintnumbers,
nofootinbib,floatfix]{revtex4-2}

\usepackage[utf8]{inputenc} 
\usepackage{amsfonts,amssymb, mathrsfs, amsmath, esint,bm,enumitem,bbold}
\allowdisplaybreaks
\usepackage{multirow}
\usepackage{subcaption}
\usepackage{ragged2e}
\DeclareCaptionJustification{justified}{\justifying}
\makeatletter    
\renewcommand\onecolumngrid{
\do@columngrid{one}{\@ne}%
\def\set@footnotewidth{\onecolumngrid}
\def\footnoterule{\kern-6pt\hrule width 1.5in\kern6pt}%
}
\renewcommand\twocolumngrid{
    \def\footnoterule{
    \dimen@\skip\footins\divide\dimen@\thr@@
    \kern-\dimen@\hrule width.5in\kern\dimen@}
    \do@columngrid{mlt}{\tw@}
}%
\makeatother 

\usepackage{diagbox}
\usepackage{latexsym}
\usepackage{graphicx}
\usepackage[dvipsnames,table]{xcolor}
\usepackage{booktabs,multirow}
\usepackage{titlesec} 
\def\thesection{\arabic{section}}
\def\thesubsection{\arabic{section}.\arabic{subsection}}
\def\thesubsubsection{\arabic{section}.\arabic{subsection}.\arabic{subsubsection}}
\usepackage{physics}
\usepackage{comment,lipsum}
\usepackage[normalem]{ulem}
\graphicspath{{./Figures/}}

\usepackage{braket}

\usepackage{tikz}
\usetikzlibrary{decorations.pathmorphing}
\usetikzlibrary{math}
\usetikzlibrary{positioning,calc}

\usepackage{hyperref}
\hypersetup{colorlinks, citecolor=ForestGreen, linkcolor=BrickRed, urlcolor=RedViolet}

\usepackage[capitalize]{cleveref}

\renewcommand{\d}{\mathrm{d}}
\newcommand{\MP}{M_\textsc{p}}

\newcommand{\TeV}{\,\text{TeV}}
\newcommand{\GeV}{\,\text{GeV}}

\newcommand{\gs}{g_\text{s}}

\def\parfrac#1#2{{\left(\frac{#1}{#2}\right)}}

\newcommand{\OGW}{\Omega_\textsc{gw}}
\newcommand{\CW}{\textsc{cw}}
\newcommand{\tree}{\text{tree}}
\renewcommand{\th}{\text{th}}
\newcommand{\tot}{\text{tot}}
\newcommand{\gst}{g_\star}
\newcommand{\SNR}{\textsc{snr}}

\begin{document}
\title{Probing Flavour Deconstruction via Primordial Gravitational Waves} 

\author{Noemi~Fabri} 
\email{noemi.fabri@physik.uzh.ch}
\affiliation{Physik-Institut, Universit\"at Z\"urich, Winterthurerstrasse 190, 8057 Z\"urich, Switzerland}
\author{Gino~Isidori} 
\email{gino.isidori@physik.uzh.ch}
\affiliation{Physik-Institut, Universit\"at Z\"urich, Winterthurerstrasse 190, 8057 Z\"urich, Switzerland}
\author{Davide~Racco} 
\email{davide.racco@physik.uzh.ch}
\affiliation{Physik-Institut, Universit\"at Z\"urich, Winterthurerstrasse 190, 8057 Z\"urich, Switzerland}
\affiliation{Institut f\"ur Theoretische Physik, ETH Z\"urich,Wolfgang-Pauli-Str.\ 27, 8093 Z\"urich, Switzerland}
\affiliation{Stanford Institute for Theoretical Physics, Stanford University, 382 Via Pueblo Mall, Stanford, CA 94305, U.S.A.}

\begin{abstract}
\noindent
We study the production of primordial gravitational waves (GWs) from first-order phase transitions (FOPTs) in extensions of the Standard Model based on Flavour Deconstruction (FD). The link fields inherent to FD generically form a rich scalar sector, with sizeable couplings at the TeV scale, providing natural conditions for strong FOPTs and correspondingly large GW emission. We identify the key parameters controlling the GW spectrum and enabling its detection at future GW observatories. 
In particular, we find that while FD scenarios can yield detectable signals, the resulting spectra typically peak at higher frequencies than the millihertz range. As a consequence, a positive observation at LISA is possible but not guaranteed,  while the signal falls in the range of mid-band proposals, making FD models  an intriguing target for upcoming GW searches. 
\end{abstract}

\preprint{ZU-TH 55/25}

\maketitle

\section{Introduction}

Primordial gravitational waves (GWs) provide a unique window into the early Universe and represent a promising novel probe of high-energy phenomena that are otherwise inaccessible to laboratory experiments. In particular, GWs produced during a cosmological first-order phase transition (FOPT) can carry indirect information about the particle content and elementary interactions occurring beyond the Standard Model (SM), at energy scales well beyond those probed directly at colliders. The next decade will witness the advent of a new generation of GW observatories. Of particular interest to this study is the LISA interferometer, designed to explore the millihertz frequency range, where cosmological signals from TeV-scale phase transitions are expected to lie~\cite{Caprini:2015zlo,Caprini:2019egz}. These experimental developments make the study of GW signatures from  motivated 
beyond-the-SM (BSM) scenarios predicting new TeV-scale dynamics especially timely and compelling.

A substantial body of work has already addressed GW production from first-order phase transitions in extensions of the SM (see~\cite{Witten:1984rs,Kosowsky:1992rz} and the comprehensive reviews in~\cite{Caprini:2018mtu,Schmitz:2020syl,Hindmarsh:2020hop,Caldwell:2022qsj,Athron:2023xlk}). Most of existing studies, however, have focused either on models inspired by grand unification (GUT-scale physics, as e.g.~Pati-Salam (PS) \cite{Huang:2020bbe}), extensions of the Higgs sector motivated by baryogenesis, and secluded sectors. By contrast, we investigate here a different class of models which has recently attracted growing attention: those based on the hypothesis of Flavour Deconstruction (FD). In FD, the observed fermion mass hierarchies are explained by the deconstruction of a gauge symmetry in flavour space, which forbids tree-level couplings among the Higgs and light-generation fermion, and where the flavour universality of the SM gauge interactions emerges as low-energy limit from family-dependent gauge groups at higher scales. Such models generically predict the existence of a rich scalar sector, consisting of multiple link fields with sizeable couplings, whose first energy layer lies around the TeV scale.  These are precisely the ingredients that can naturally lead to strong FOPTs, and consequently to potentially observable GW signals in mHz--Hz frequency range.

The GW phenomenology of FD has only begun to be explored. To date, the only dedicated study is the pioneer analysis of~\cite{Greljo:2019xan},
which considered a specific realisation of the FD hypothesis: the so-called  ${\rm PS}^3$ model~\cite{Bordone:2017bld}. The possibility of revealing the threefold peak structure associated to the three scales of the  ${\rm PS}^3$ model, discussed in~\cite{Greljo:2019xan}, can be considered the 
most optimistic scenario to reveal an underlying FD structure. However, it might not be representative of a broader class of FD models. Motivated by the increasing interest in such models~\cite{Bordone:2017bld,Greljo:2018tuh, Fuentes-Martin:2020bnh, Allwicher:2020esa, Fuentes-Martin:2020hvc,Fuentes-Martin:2022xnb, 
FernandezNavarro:2023rhv, Davighi:2023iks, Davighi:2023evx,  Davighi:2023xqn, Barbieri:2023qpf, Fuentes-Martin:2024fpx, FernandezNavarro:2024hnv, Greljo:2024ovt, Covone:2024elw, Davighi:2025cqx, Lizana:2024jby, FernandezNavarro:2025zmb}  and by the recognition that they span a large and varied parameter space~\cite{Davighi:2023iks}, we revisit the problem in more general terms.
Our primary goal is to identify the key parameters and conditions that control the strength of the FOPT in FD models, and thus determine the resulting GW spectra. In doing so, we aim to establish under which circumstances these models can give rise to signals within the sensitivity reach of future space-based detectors. We concentrate in particular on the case of TeV-scale phase transitions, which arise naturally in FD setups, and which are especially relevant for LISA. This focus not only allows us to connect theoretical predictions with near-future observational prospects, but also highlights the broader phenomenological role of FD as a framework that links flavour physics to cosmological observables.

The paper is organised as follows: in Section~\ref{sect:model}
we introduce the model setup, identifying the key parameters relevant to describe the FOPT. The dynamics of the PT and the ingredients necessary to derive the GW spectrum are discussed in Section~\ref{sect:PT}. The results for the GW spectra are discussed in Section~\ref{sect:Disc}. The main findings are summarised in the Conclusions.

\section{The model setup}
\label{sect:model}

As prototype PT around TeV scale, motivated by the FD hypothesis, we consider the one associated to the following spontaneous symmetry breaking (SSB): 
\begin{equation}
     SU(4)^{[3]}\times SU(3)^{[12]} \times U(1)' \to 
     SU(3)_c \times U(1)_Y\,.
     \label{eq:4321}
\end{equation}
The transition is assumed to be generated by an appropriate set of scalar fields (so-called link fields) acquiring a non-trivial vacuum expectation value (VEV) at a scale $v$.
In Eq.~(\ref{eq:4321}) the superscript denotes flavour non-universal gauge groups acting only on third or light fermions families, while $SU(3)_c$ and $U(1)_Y$ are the flavour-universal SM gauge groups associated to colour 
and hypercharge.\footnote{For the two lightest fermion families, the charges under the flavour non-universal $U(1)^\prime$ group coincide with the SM hypercharge. The $U(1)^\prime$ charges of third-generation fermions are non-vanishing only in the right-handed case. Denoting $Q_R^{\prime[3]}$ such charges,  third-generation hypercharge can be written as $Y^{[3]}=\sqrt\frac{2}{3} T_{15}^{[3]} + Q_R^{\prime[3]}$, where $T^{[3]}_{15} = \sqrt{\frac{3}{8}} (B-L)^{[3]}$ 
is the colour-singlet generator of $SU(4)^{[3]}$.}
This transition  can be viewed as the last step of the SSB chain that connects a completely  flavour non-universal gauge group in the ultraviolet (UV)  down to the SM gauge group ($G_{\rm SM}$), as it happens for instance in the ${\rm PS}^3$ model~\cite{Bordone:2017bld}. As pointed out in~\cite{Davighi:2023iks}, the step in Eq.~(\ref{eq:4321})
is an unavoidable expectation of models featuring a semi-simple embedding in the UV. More generally, we argue that this choice represents a generic benchmark for a broad class of FD models characterised by a large multiplicity of broken generators around the TeV scale. 

The SSB transition occurring at the TeV scale is the most constrained of the possible steps defining the SSB chain of the complete model: landing to $G_{\rm SM}$ implies precise matching conditions for the gauge couplings at the SSB scale. The value of $v$ is expected to be in the few TeV range in order to satisfy the criterion of finite naturalness for the Higgs mass~\cite{Allwicher:2020esa,Davighi:2023iks}, with the lowest phenomenologically allowed values being theoretically favoured. 
Collider constraints on the new gauge bosons require $v\gtrsim 1\TeV$ (assuming 
$g_4 \gtrsim 1$ for the $SU(4)$ gauge coupling)~\cite{Haisch:2022afh,Aebischer:2022oqe}. In the following we choose  $v= 1\TeV$  and $v= 3\TeV$
as  reference values for the numerical analysis.

For a quantitative description of the PT, we need to specify the following ingredients from the full theory:
\begin{enumerate}
\item[(\textit{i})] The structure of the scalar potential for the scalar fields acquiring a non-trival VEV. To simplify our analysis, we identify a region of parameter space where the VEV is 
acquired by a single scalar field ($\phi$), whose mass is significantly lighter with respect to those of the other scalar degrees of freedom. The latter can effectively be decoupled at energy scales above $v$. Under these hypotheses, the tunnelling reduces to a one-dimensional problem, fully described by the dynamics of~$\phi$.
\item[(\textit{ii})] The value of the gauge couplings at the SSB scale, which determine the impact of vector bosons on the thermal effective potential. 
According to the FD hypothesis and collider constraints, a natural range for 
the $SU(4)$ coupling is $g_4 (v) \gtrsim 1$. Once  $g_4$ is fixed, the other couplings of the gauge group before SSB are fixed in terms of SM gauge couplings  by the matching conditions. In particular, the following relation holds
\begin{equation}
g_4^{-2}(v) +g_3^{-2}(v)= \gs^{-2}(v)\,,
\label{eq:matching}
\end{equation}
where $\gs$ is the QCD coupling. 

\item[(\textit{iii})] The Yukawa couplings of the fermions interacting with $\phi$. 
In general, these can affect the thermal potential. However, in our setup these fermions have large 
vector-like masses and  Yukawa couplings smaller than unity~\cite{Davighi:2023iks}, hence they play a 
negligible in the PT. 
\end{enumerate}

For concreteness, we specify the model as follows.
We assume that the SSB in Eq.~(\ref{eq:4321}) is due to the VEV of two fields $\Omega_3\sim (\mathbf{\overline 4},\mathbf{3},\mathbf{1},\tfrac 16)$ and $\Omega_1\sim (\mathbf{\overline 4},\mathbf{1},\mathbf{1}, -\tfrac 12)$~\cite{Fuentes-Martin:2020hvc,DiLuzio:2018zxy}. To minimise the free parameters in the scalar sector we further assume the scalar potential respects a global 
 $SU(4)^{[3]} \times SU(4)^{[12]}$  custodial symmetry~\cite{Fuentes-Martin:2020hvc,Houtz:2022fnk}, 
 with $\Omega_3$ and  $\Omega_1$ embedded into a single   $\Omega_4$ field transforming as a $\bar 4\times 4$  of $SU(4)^{[3]} \times SU(4)^{[12]}$. 
 Under these assumptions, the scalar potential assumes the form
\begin{multline}
\label{eq:Omega4 Lagr}
V=  \mu^2 \Tr(\Omega_4^\dagger\Omega_4)
    +\rho_1 \left(\Tr (\Omega_4^\dagger\Omega_4)-v^2\right)^2 \\
    + \rho_2 \Tr \left((\Omega_4^\dagger\Omega_4) -\tfrac 12 v^2 \mathbb{1}_{4}\right)^2\\
    + \rho_3 \epsilon_{\alpha \beta \gamma \delta}\epsilon^{\rho \sigma \mu \nu}(\Omega_4)_\rho^\alpha(\Omega_4)_\sigma^\beta(\Omega_4)_\mu^\gamma (\Omega_4)_\nu^\delta+ \text{h.~c.}
\end{multline}

To understand under which conditions the scenario described in (\textit{i}) is realised, it  is useful to decompose $\Omega_4$ under SM representations,
\begin{equation}
\Omega_4=\begin{pmatrix}
\frac{1}{\sqrt{2}}v I_{3}+\frac{1}{\sqrt{6}}S_3 I_3+ O_a t^a & T_1^* \\
T_3 &\frac{1}{\sqrt{2}}v+S_1
\end{pmatrix}\,.
\end{equation}
Here  $S_{1,3} \sim (1,1)_0$  are SM singlets, $O_3 \sim (8,1)_0$  is an  $SU(3)_c$ octet,
and $T_{1,3}\sim (3,1)_{2/3}$  are $SU(2)_L$ triplets.
The one-dimensional scenario is realised when the radial component of $\Omega_4$ exhibits small  mixings with the other states and
its massive excitation,
$S_\phi \propto \sqrt{3}(S_3+S_3^*)+ \sqrt{2}(S_1+S_1^*)$,
has a small mass.
This occurs for $\rho_{1,2,3}\sim \mathcal O(1)$ in the limit 
$| 12\rho_3 -\rho_2 |\ll 1$ and $0 < 6\rho_3+ \rho_1  \ll 1$.%
\footnote{The other SM-singlet eigenstates describe a massless Goldstone boson and two massive scalars. Large positive mass terms for the latter 
are realised for $\rho_3 >0$, and adding 
a soft custodial-breaking term to raise the mass of 
$\tilde \phi \propto \sqrt{2}(S_3+S_3^*)- \sqrt{3}(S_1+S_1^*)$.}
In this limit $\phi = {\rm Tr}(\Omega_4)/(2\sqrt{2})$ and 
$\lambda = \tfrac 52 (6\rho_3+ \rho_1) \ll 1$.
To a good accuracy, the part of the potential depending on $\phi$ assumes the form
\begin{equation}
\label{eq:V tree}
V_\text{tree}(\phi) =  \frac{\lambda}{4}\phi^4 -\frac 12 \lambda v^2 \phi^2 \,
\end{equation}
and the free parameters  characterising the PT reduce to the two effective couplings $\lambda(v)$ and $g_4(v)$.
 
The 15 broken generators in Eq.~(\ref{eq:4321}) are in one-to-one correspondence with the field-dependent masses of the massive gauge bosons, namely the coloron ($G$), the vector leptoquark ($U$) and the $Z'$ boson, collectively denoted $V \equiv \{G,U,Z'\}$. The tree-level expressions for their masses can be written as $m_V = g_V v$, where 
\begin{equation}
\label{eq:g bosons}
g_G^2 = \frac 12 \left(g_4^2+g_3^2\right) \,, \ 
g_U^2 = \frac 12 g_4^2\,, \ 
g_{Z'}^2 = \frac 12 g_4^2+\frac 13 g_1^2\,,
\end{equation}
with corresponding multiplicities  (product of three polarisation states times electric and colour charges) 
given by
\begin{equation}
\label{eq:c bosons}
c_G = 24 \,, \quad
c_U = 18\,, \quad
c_{Z'} = 3 \,.
\end{equation}

 \medskip

While the SSB transition in Eq.~(\ref{eq:4321})
provides a good benchmark for all models with a large multiplicity of broken generator at the TeV scale, it is not representative of the minimalistic FD models presented in~\cite{Davighi:2023evx,FernandezNavarro:2023rhv,Barbieri:2023qpf}, which involve only $U(1)$ groups. 
To this purpose, we consider an alternative benchmark where the last SSB step before reaching $G_{\rm SM}$ is
\begin{equation}
     U(1)^{[3]}_{B-L} \times U(1)' \to  U(1)_Y\,.
     \label{eq:111}
\end{equation}
 We treat this case in full analogy with the main benchmark discussed above. 
To facilitate the comparison among the two cases, we denote $g_4$ the coupling of the $U(1)^{[3]}_{B-L}$ gauge group using the same normalization of the charges as for the $T^{[3]}_{15}$ generator of $SU(4)^{[3]}_{15}$.
We further denote $\lambda$ the quartic coupling of the scalar field acquiring VEV, such that ($\lambda$, $g_4$)
still denote the key model parameters. 
In this case we have a single broken generator, 
$V \equiv \{Z'\}$, with $c_{Z'} = 1$ and 
$m_{Z'}^2 = v \sqrt{g_4^2/2+g_1^2/3}$, and the 
 matching condition reads~$\tfrac 23 g_4^{-2}(v) +g_1^{-2}(v)= g_Y^{-2}(v)$.

\section{Analysis of the phase transition}
\label{sect:PT}

In this section, we review our treatment of the PT. 
Thermal PTs have been studied for a long time, dating back to the seminal papers on thermal symmetry restoration and the thermal effective potential \cite{Dolan:1973qd,Linde:1981zj,Anderson:1991zb}. 
One of the most relevant phenomenological motivations, has been the study of the metastability of our current Higgs vacuum in the SM \cite{Cabibbo:1979ay,Altarelli:1994rb,Espinosa:2007qp,Ellis:2009tp,Elias-Miro:2011sqh, Degrassi:2012ry, Buttazzo:2013uya,Espinosa:2015qea}.

Our purpose is to highlight the main regions of the parameter space $(\lambda,g_4)$ where the GWB generated by the PT can be loud enough to be detectable. 
Here  $\lambda$, defined in Eq.~\eqref{eq:V tree}, denotes the tree-level effective quartic coupling of the lightest singlet field $\phi$ controlling the phase transition.

In the FD models we consider, the contribution of fermions to the thermal potential of $\phi$ is negligible with respect to the vector contribution in this class of models.
We use then the expression for the thermal effective potential obtained at one loop in perturbation theory \cite{Anderson:1991zb}
\begin{equation}
\label{eq:V thermal}
V_\th (\phi,T) = \sum_{V} c_V T^4 J_B\left( g_V^2 \frac{\phi^2}{T^2}\right),
\end{equation}
where the function $J_B(\alpha)$, offset so that it vanishes at the origin, reads
\begin{equation}
\label{eq:JB}
J_B(\alpha) = \frac{\pi^2}{90} + \frac{1}{2\pi^2}\int x^2 \ln \left( 1-e^{-\sqrt{x^2+\alpha}} \right) \d x\,.
\end{equation}
We compute $J_B$ numerically, without resorting to a high-temperature analytic expansion that is not valid for the typical temperatures $T_n\sim (0.1-0.5)v$ and couplings $g\sim \mathcal O(1)$ that we consider. 
As we do not perform a complete treatment of the PT at two loops in perturbation theory, we do not account for daisy resummation (which can affect the effective potential for $\phi\lesssim T_n$ around the PT%
\footnote{Daisy resummation is likely to have  a non-negligible impact on our predictions; however,  we expect it not to change the overall picture. 
Fig.~\ref{fig:U1case} shows that $T_n \simeq 0.3 v$ for the main model we consider, and $\phi\sim T_n$ lies just around the barrier location from Fig.~\ref{fig:4321_potential}, hence the impact of daisy resummation on the barrier should be moderate.}).
Including such effects, and computing the dimensionally-reduced effective potential, would be relevant to study the PT more systematically.

The one-loop effective potential at zero temperature is given by the Coleman-Weinberg (CW) potential \cite{Coleman:1973jx}
\begin{equation}
\label{eq:V CW}
V_\CW (\phi)= \frac{1}{64\pi^2} \sum_{V}
   c_V g_V^4 \phi^4 \left(\ln \left( \frac{g_V^2 \phi^2}{\Lambda^2} \right) -\frac 56 \right)
\end{equation}
where we take the RG scale $\Lambda=v$, and we omit the fermions as in the class of models we consider their effect is subleading to vector bosons.
As we explain later, the CW contribution to the potential plays an important role in the parameter space that is most relevant for PTs.
For notational convenience, we define
\begin{equation}
V_\tot(\phi,T) = V_\tree(\phi)+V_\CW(\phi)+V_\th(\phi,T) \,.
\end{equation}

Having defined the effective potential with thermal corrections, we can compute the nucleation temperature at which the light scalar $\phi$ tunnels from the metastable vacuum in the origin to the true vacuum $\phi_\text{min}(T)\approx v$ ($V_\CW(\phi)$ changes the position of the minimum also at $T=0$).
The problem of tunnelling of a quantum field from a metastable minimum has been studied for a long time, starting from the  semiclassical treatment of the bounce \cite{Coleman:1977py} (see also \cite{Espinosa:2018hue,Espinosa:2018voj} for a recent revisitation of the approach). 
In practice, the numerical calculation of the bounce is not an easy task, and various numerical packages have been developed (see e.g.~\cite{Wainwright:2011kj,Masoumi:2016wot,Guada:2020xnz,Athron:2024xrh}, or the recent \cite{Brdar:2025gyo} which interfaces with  \cite{Ekstedt:2022bff} to compute the dimensionally reduced effective potential).
We use here \texttt{AnyBubble} \cite{Masoumi:2016wot} to compute the tunnelling action $S(T)\equiv S_3(T)/T$ as a function of the temperature $T$. 
We perform a standard perturbative study at one loop of the effective potential, although higher-order effects are likely to have an impact (see \cite{Chala:2024xll,Lewicki:2024xan,Ekstedt:2024etx,Bernardo:2025vkz} for recent studies).

The next complex task in the study of a FOPT involves the thermodynamics of the bubble walls and the plasma coupled to the scalar field \cite{Ignatius:1993qn,Moore:1995si,Espinosa:2010hh}. 
For simplicity, we calculate the thermodynamical variables that characterise the FOPT (the nucleation temperature $T_n$, the transition strength $\alpha$ and the inverse of the PT time duration $\beta$) as \cite{Caprini:2015zlo}
\begin{align}
\label{eq:Tn}
T_n & : \quad 
\Gamma \approx T^4 e^{-S(T)} \overset{!}{=}H(T)^4\\
\label{eq:alpha}
\alpha & : \quad 
\alpha =\left[
\frac{V_\tot(0,T)-V_\tot(\phi_\text{min}(T),T)}{\rho_r(T)}
\right]_{T=T_n}\\
\label{eq:beta}
\beta & : \quad 
\frac{\beta}{H(T)}=\left. T \frac{\mathrm d S(T)}{\mathrm d T}\right|_{T=T_n}
\end{align}
where $\rho_r(T) = \tfrac{\pi^2}{30}\gst T^4$, $H^2(T) \approx \rho_r(T)/(3\MP^2)$, $\MP$ is the reduced Planck mass, and we take $\gst(T_n) = 200$ as a representative value.
A more sophisticated calculation of $T_n$ would include prefactors in the evaluation of the functional determinant \cite{Callan:1977pt,Coleman:1987rm,Strumia:1998nf,Isidori:2001bm, Dunne:2005rt} (recently revisited and refined in \cite{Ai:2023yce,Baratella:2024hju,Carosi:2024lop,Baratella:2025dum}) in \cref{eq:Tn}, which affect though $T_n$ only logarithmically.

We also set aside the calculation of the net friction experienced by the bubble walls and the corresponding wall velocity $v_w$ \cite{DeCurtis:2022hlx,Laurent:2022jrs,Ai:2023see,DeCurtis:2024hvh,Krajewski:2024gma,Ai:2025bjw,Carena:2025flp} (see e.g.~\cite{Ekstedt:2024fyq} for a numerical package), setting for simplicity $v_w=1$. 
In reality, we do not expect runaway wall dynamics, due to the large couplings between gauge bosons and Higgs field.
A realistic value of $v_w$ would imply a small reduction of the amplitude of the GWB, and would be an ingredient of an accurate prediction of the signal.

We can finally relate these quantities to the GW energy fraction $\OGW h^2$, that for ease of comparison we take from the LISA Cosmology white paper \cite{Caprini:2015zlo}:
\begin{subequations}
\label{eq:OGW}
\begin{gather}
h^2 \OGW \approx h^2 \Omega_\text{sw} + h^2 \Omega_\text{turb} \,,\\
h^2 \Omega_\text{sw} = 2.65\cdot 10^{-6} \frac{H_n}{\beta} \parfrac{\kappa_\text{sw} \alpha}{1+\alpha}^2 \parfrac{100}{\gst}^{\frac{1}{3}} v_w S_\text{sw}(f)\,, \\
S_\text{sw}(f) = \parfrac{f}{f_\text{sw}}^3 \parfrac{7}{4+3(f/f_\text{sw})^2}^{\frac 72}\,,\\
\kappa_\text{sw} = \frac{\alpha}{0.73+0.083\sqrt{\alpha}+\alpha} \,,\\
\label{eq:f pk GW}
f_\text{sw} = 1.9 \cdot 10^{-5}\text{\,Hz} \cdot \frac{1}{v_w} \frac{\beta}{H_n} \frac{T_n}{100\GeV} \parfrac{\gst}{100}^{\frac 16}\,,\\
h^2 \Omega_\text{turb} = 3.35\cdot 10^{-4} \frac{H_n}{\beta} \parfrac{\kappa_\text{turb} \alpha}{1+\alpha}^{\frac 32} \parfrac{100}{\gst}^{\frac{1}{3}} v_w S_\text{turb}(f)\,, \\
S_\text{turb}(f) = \parfrac{f}{f_\text{turb}}^3 \parfrac{1}{1+f/f_\text{turb}}^{\frac{11}{3}} \frac{1}{1+8\pi f/h_n}\,,\\
h_n =  1.65 \cdot 10^{-5}\text{\,Hz} \cdot  \frac{T_n}{100\GeV} \parfrac{\gst}{100}^{\frac 16}\,,\\
\kappa_\text{turb} = 0.05 \,\kappa_\text{sw} \,,\\
f_\text{turb} = 1.4 f_\text{sw}\,,
\end{gather}
\end{subequations}
where the subscripts $_\text{sw}$ and $_\text{turb}$ refer to the contributions from sound waves and plasma turbulences. 
We neglect the contribution from the bubble walls, which is typically subdominant if they reach a terminal velocity due to the friction of the plasma.

Finally, we can use the GW spectrum defined in \cref{eq:OGW} to compute the signal-to-noise ratio (SNR) for the LISA experiment, as
\begin{equation}
\SNR = \sqrt{T_\textsc{lisa}\int \mathrm d f \parfrac{h^2 \OGW(f)}{h^2 \Omega_\text{noise}(f)}^2} \,,
\end{equation}
where the integral is carried out in the whole frequency range where LISA is sensitive, with a noise spectrum $h^2\Omega_\text{noise}(f)$ as given in the blue curve of Fig.~11 of \cite{Babak:2021mhe}, and $T_\textsc{lisa}=4 y$.

\begin{figure}[t]
\centering
\includegraphics[width=\columnwidth]{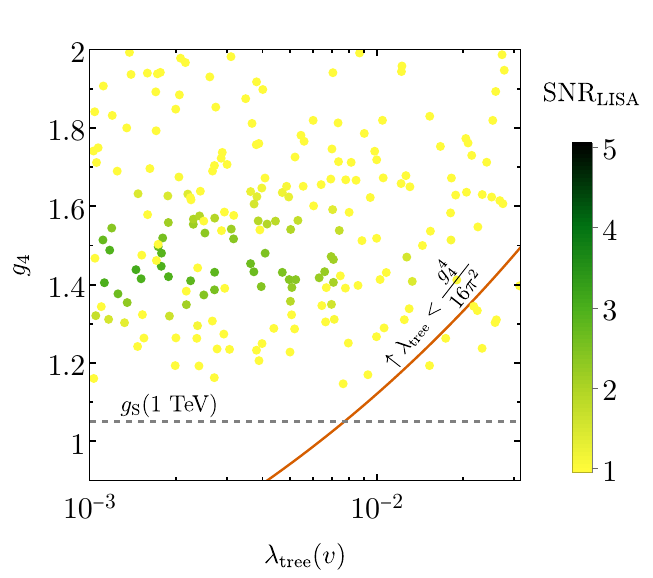}
\caption{SNR at LISA for the GW emission induced by the  
SSB transition $SU(4)^{[3]}\times SU(3)^{[12]} \times U(1)' \to 
     SU(3)_c \times U(1)_Y$,  in the $g_4$--$\lambda$ plane, assuming $v=1~\TeV$. }
\label{fig:4321_l_g4}
\end{figure}

\section{Discussion}
\label{sect:Disc}

In Fig.~\ref{fig:4321_l_g4} we show the expected SNR at LISA in the $g_4$--$\lambda$ plane, for $v=1~\TeV$,
in the benchmark scenario defined by  Eq.~\eqref{eq:4321}.  
As can be seen, achieving a sizeable SNR requires values of $g_4$ of order unity and relatively small quartic couplings $\lambda$. In most of the parameter space that leads to a strong FOPT, $\lambda$ is dominated by the one-loop contribution rather than by the tree-level term, as indicated by the red line in Fig.~\ref{fig:4321_l_g4}.  

\begin{figure*}[t]
\centering
\includegraphics[width=.92\columnwidth]{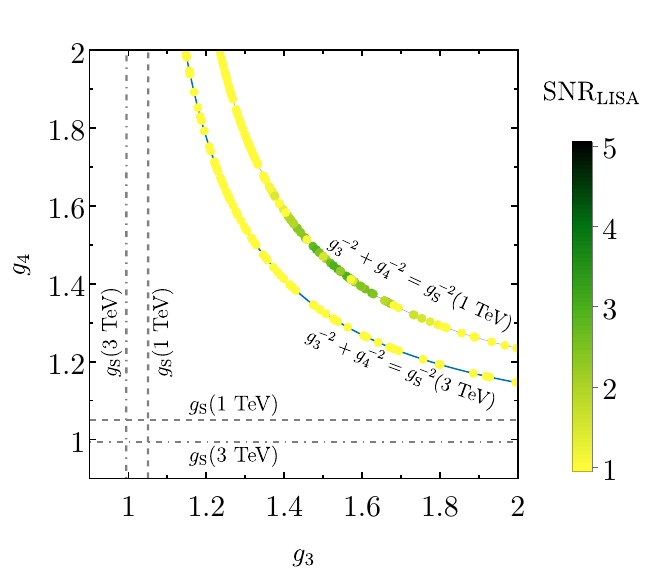} 
\hspace*{1em}
\includegraphics[width=\columnwidth]{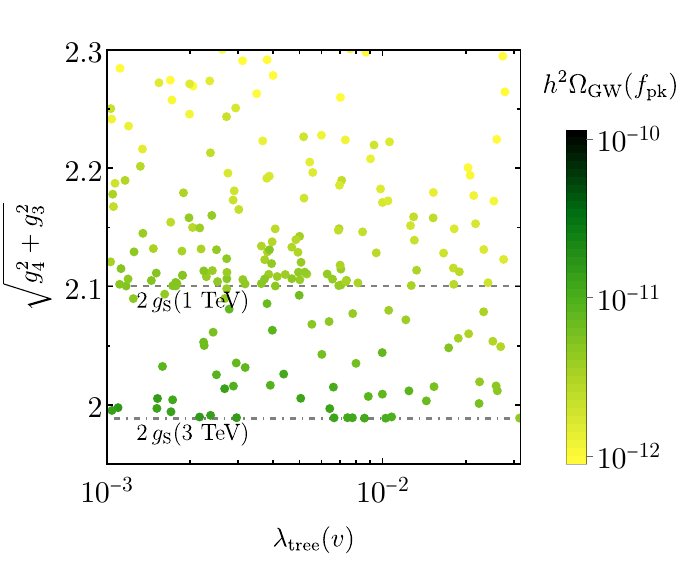}
\caption{\textit{Left panel}: SNR at LISA for the 
SSB transition $SU(4)^{[3]}\times SU(3)^{[12]} \times U(1)' \to  SU(3)_c \times U(1)_Y$,  in the $g_4$--$g_3$ plane, assuming $v=1~\TeV$ (as in Fig.~\ref{fig:4321_l_g4}).  
\textit{Right panel}: $\Omega_{\rm GW}$ for the same FOPTs in the $\sqrt{g_4^2+g_3^2}$--$\lambda$ plane; the horizontal lines denote the minimal value of  $\sqrt{g_4^2+g_3^2}$
set by the matching condition~(\ref{eq:matching}) for 
 $v=1~\TeV$ and  $v=3~\TeV$.}
\label{fig:4321_g3g4}
\end{figure*}

The restricted range of viable $g_4$ values, clustered around $g_4 \approx 1.5$, is further clarified in Fig.~\ref{fig:4321_g3g4}. Rather than $g_4$ alone, the key parameter controlling the efficiency of the GW production is the combination $\sqrt{g_3^2 + g_4^2}$. This behaviour is expected, since the mass of the coloron---the heavy vector boson that, due to its large multiplicity, dominates the thermal effective potential---is proportional to $\sqrt{g_4^2+g_3^2}$. The right panel of Fig.~\ref{fig:4321_g3g4} shows that smaller values of $\sqrt{g_4^2+g_3^2}$ lead to stronger GW signals. However, $g_3$ and $g_4$ are not independent parameters, but are related through the matching condition in Eq.~\eqref{eq:matching}. As a result, the minimum achievable value of $\sqrt{g_3^2 + g_4^2}$ occurs for $g_4 \approx 1.5$, as illustrated in the left panel of Fig.~\ref{fig:4321_g3g4}.

\begin{figure*}[t]\centering
\includegraphics[width=1.5\columnwidth]{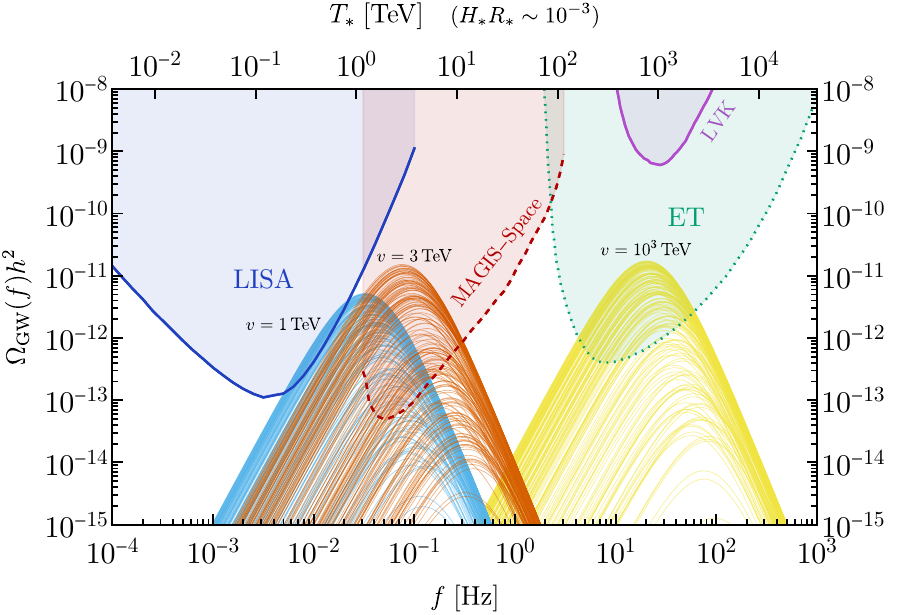}
\caption{GW spectrum induced by the SSB transition $SU(4)^{[3]}\times SU(3)^{[12]} \times U(1)' \to  SU(3)_c \times U(1)_Y$ for different values of $g_4$ and $\lambda$, compared to the expected sensitivities of future GW observatories. The different colors correspond to different choice of $v$: 
1~TeV (blue), 3~TeV (brown), and $10^3$~TeV (yellow).
The latter value is shown for illustrative purposes only.}
\label{fig:4321_GW_spectra}
\end{figure*}

An additional point to highlight is the dependence on the 
symmetry-breaking scale $v$. Because of the running of $g_s$, the relation between $g_3$ and $g_4$ changes with $v$. Larger values of $v$ correspond to smaller $g_s(v)$, and consequently smaller $\sqrt{g_3^2 + g_4^2}$, whose minimum value is $2g_s(v)$.
This is why $\Omega_{\rm GW}$ is larger for $v=3$~TeV with respect to $v=1$~TeV 
(Fig.~\ref{fig:4321_g3g4}, right panel). 
At the same time, increasing $v$ shifts the peak frequency of the GW spectrum to higher values, moving the signal away from the maximum sensitivity of LISA. This behaviour is clearly illustrated in Fig.~\ref{fig:4321_GW_spectra}, where we display the GW spectra for different choices of $v$.

\begin{figure*}[t]\centering
\includegraphics[width=\textwidth]{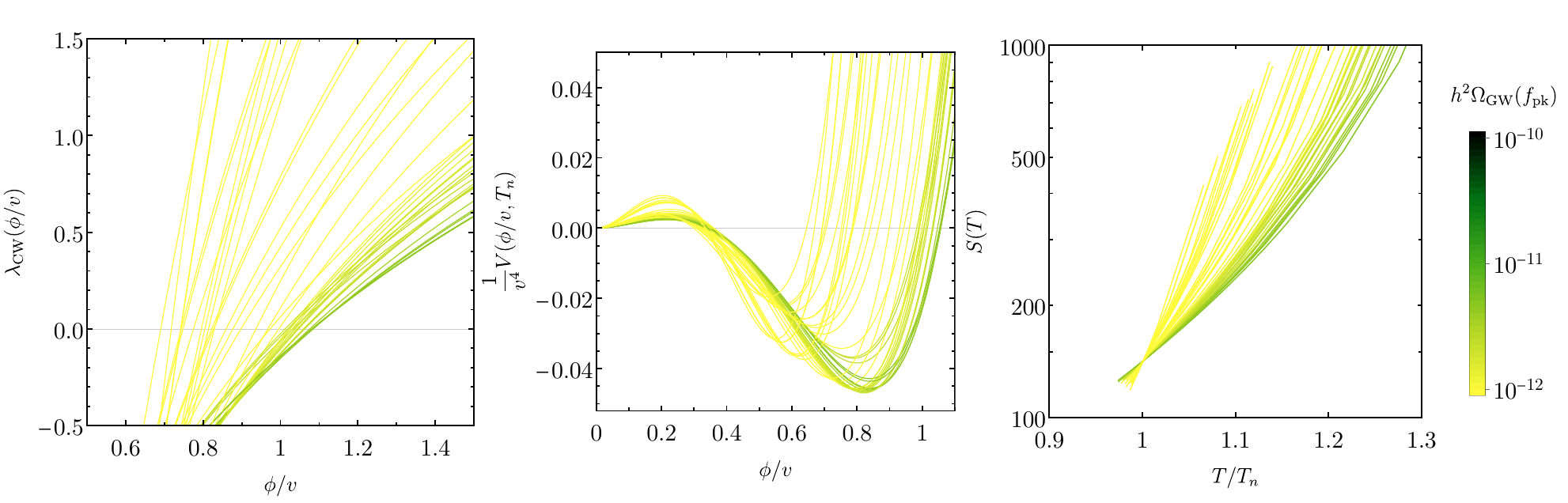}
\caption{\textit{Left panel}: running of the quartic coupling of $V_\tree(\phi)+V_\CW(\phi)$ for the same points of \cref{fig:4321_l_g4}. In all these panels, the lines are colour-coded according to the amplitude of the $\OGW$ signal.
\textit{Central panel}: shape of the effective potential $V_\tot(\phi,T_n)$ evaluated at the nucleation temperature for each point of the scan, showing the potential energy difference (proportional to $\alpha$).
\textit{Right panel}: temperature dependence of the tunnelling action when approaching $T_n$ for each point of the scan, showing that a slower dependence on $T$ (corresponding to smaller $\beta$) implies a stronger $\OGW$.}
\label{fig:4321_potential}
\end{figure*}

\begin{figure*}\centering
\includegraphics[width=\textwidth]{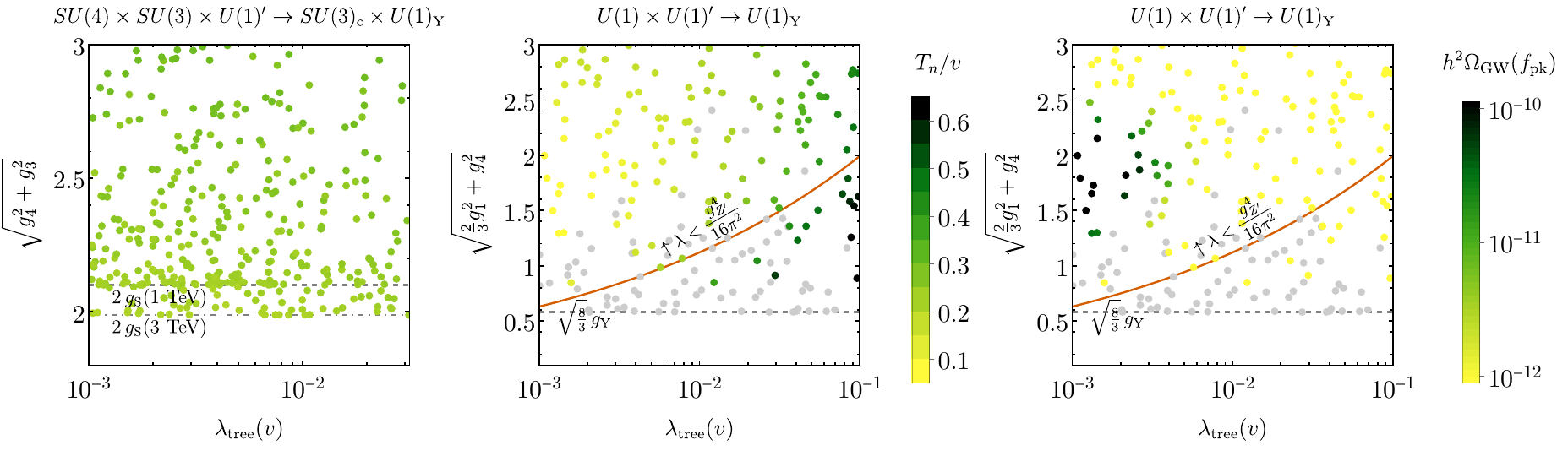}
\caption{\label{fig:U1case}
Comparison of results obtained assuming different 
SSB patterns.
\textit{Left and central panel}: scan of the nucleation temperature $T_n/v$ across the parameter space of $\lambda_\tree$ and the combination of gauge couplings controlling the mass of the most relevant massive gauge boson. 
The left panel refers to the SSB pattern in \cref{eq:4321}, considered in all previous figures, while the central and right panels refer to the $U(1)$ model in \cref{eq:111}.
The grey-coloured points correspond to model parameters of the $U(1)$ model where, due to numerical instabilities, the solver is unable to identify the narrower range for the nucleation, so we cannot identify the FOPT. 
\textit{Right panel}: same as the central panel, showing the amplitude of the GW background.
In all plots, the red lines mark where the effective quartic coupling is dominated by one-loop contributions rather than tree-level.}
\end{figure*}

For illustrative purposes, in addition to the natural choices 
$v=1$  and 3 TeV, in Fig.~\ref{fig:4321_GW_spectra} we also show the spectrum obtained for $v=10^3$~TeV. As can be seen, for 
phase transitions occurring around that high energy scale the peak frequency falls in the sensitivity window of third-generation ground-based interferometers such as ET and CE. In this regime the prospects for detection are promising, although the corresponding model-building realisations are subject to greater theoretical uncertainties.  This result is consistent with the findings of Ref.~\cite{Greljo:2019xan}.

To understand why lower values of $\sqrt{g_3^2 + g_4^2}$ are preferred, we need to look at the shape of the effective CW potential generated beyond the tree level, 
as in \cref{eq:V CW}. 
The GW strength of the transition is maximised for larger $\alpha$, corresponding to a larger latent heat injected in the plasma through the bubble dynamics, and for smaller $\beta$, implying that bubbles nucleate less frequently and more sparsely, so that they are larger when plasma shock waves collide.  
From a microscopical perspective, both properties are achieved when $V_\tree(\phi)+V_\CW(\phi)$ has a slower variation in $\phi$ between the origin and the minimum (which roughly implies a larger $\alpha$), so that the thermal corrections $V_\th(\phi,T)$ around $T_n$ are small (generically implying a smaller $\beta$).
This intuition is confirmed by Fig.~\ref{fig:4321_potential}.
For large $\sqrt{g_3^2 + g_4^2}$ the running of $\lambda$ is too fast, as shown in the left panel. 
This reduces the latent heat (potential energy difference, see the central panel) and the duration of the transition (as the potential evolves faster with $T$ and $\phi$, see right panel), because of the shift to smaller $\phi$ of the minimum.
Notice that, when quantum corrections are included, the minimum of the potential is not exactly in $v$, as in the tree-level case. The peak frequency of the GW spectrum is related to $T_n$ in \cref{eq:f pk GW}, rather than the minimum location.

\medskip 

To conclude our analysis, in Fig.~\ref{fig:U1case} we present results that illustrate the main modifications occurring  if we move from the (non-Abelian) SSB pattern in Eq.~\eqref{eq:4321} ---adopted in all previous figures--- to the $U(1)$ model in Eq.~\eqref{eq:111}. A clear comparison of the two cases is obtained by looking at the right panel in Fig.~\ref{fig:4321_g3g4} 
vs.~the most right panel in Fig.~\ref{fig:U1case}.
As can be seen, large $\OGW$ values can occur also in the $U(1)$ case; however, only for very small values of $\lambda$.  
Due to the smaller multiplicity, in the Abelian case a sufficiently strong FOPT occurs only via a delicate (and numerically unstable) tuning between tree-  and loop-corrections to the Higgs potential.  This conclusions can be deduced by looking at the central and left panels in Fig.~\ref{fig:U1case}, where we compare the $T_n/v$ ratio in the two cases: while this is almost constant over the phase space in the non-Abelian model,  it changes rapidly in the $U(1)$ case. In the latter case a large GW emission occurs for $T_n/v$ as low as 0.1, corresponding to large quantum corrections (rapidly varying with the model parameters) vs.~the tree-level contributions to the scalar potential.

From these considerations, we deduce that it is less likely to observe a GW signal in the $
U(1)$ case. Nevertheless, the existence of parameter regions yielding observable signals in both Abelian and non-Abelian scenarios highlights the challenge of disentangling the underlying theory in the event that a GW signal were detected, relying solely on GW data. An attractive feature of this class of models, however, is that they predict a FOPT around the TeV scale, where complementary evidence could be obtained at colliders, both from direct searches and precision measurements.

\section{Conclusions}
\label{sect:Conc}

We have explored the GW phenomenology of flavour deconstruction  models, focusing on the spontaneous symmetry-breaking steps that naturally occur, in such models, around the TeV scale. Using two general benchmarks for the underlying SSB transition, we have shown that the key parameters controlling the strength of the phase transition are essentially two combination of parameters: an effective combination of quartic scalar couplings 
and a effective combination of gauge couplings. 
Confirming previous studies, we have shown that sizeable GW signals can be generated in regions of the parameter space where  loop corrections (controlled by the gauge couplings) compete with respect to tree-level 
contributions (controlled by the quartic coupling) to the 
scalar potential. We have also outlined the important role of the matching conditions among the gauge couplings, a point that was not stressed in previous studies: these are particularly relevant for TeV-scale transitions, where the gauge group after SSB is the SM one, hence the structure of the theory is highly constrained.

In the non-Abelian benchmark, characterised by a high multiplicity of broken generators, natural regions of the parameter space lead to strong first-order transitions with potentially detectable signals at LISA. These regions are characterised by $O(1)$ values for the  flavour non-universal gauge coupling characterising third-generation dynamics,  
which are favoured by the FD hypothesis. The Abelian benchmark, by contrast, yields observable signals only in finely tuned corners of parameter space, making GW detection less likely.

Compared to the pioneer study of Ref.~\cite{Greljo:2019xan}, we confirm that FD models can indeed give rise to observable GW signals, possibly with the emergence of multiple peaks. 
However, our results indicate that the peaks in the spectrum might not be all clearly detectable, as the production of a strong GW background is not generic. 
In particular, the TeV-scale peak is strongly affected by the matching conditions among gauge couplings and by the limited sensitivity of LISA in the relevant frequency window: the peak of the GW emission tend to be a frequencies (slightly) above the millihertz range. 
On the other hand, it is worth stressing that our study of the sensitivity suffers of various limitations: more realistic studies in this frequency window would be very welcome. 

A important outcome of our study is that both the Abelian and the non-Abelian benchmarks can in principle generate detectable GW signals. 
Not surprisingly, in case of a signal detection, GW data alone would lead to a large degeneracy not only in the parameters of the underlying model, but also in model space.  
However, the TeV-scale dynamics and large couplings inherent in the FD hypothesis implies that complementary evidence on such class of models is expected also 
from collider experiments, especially via indirect searches.  
Collider experiments and GW observatories would probe different combinations of effective couplings, hence their combined information 
would  provide a powerful tool to achieve a clear identification of the underlying theory.

\begin{acknowledgments}
We thank Admir~Greljo and Lucio~Mayer for helpful comments and discussions.
We acknowledge use of the publicly available code {\tt AnyBubble}~\cite{Masoumi:2016wot}.
This project has received funding by the Swiss National Science Foundation~(SNF) under contract~200020-204428.
D.R.~is supported at U.~of Zurich by the UZH Postdoc Grant 2023 Nr.\,FK-23-130, and was supported at Stanford U.\,by NSF Grant PHY-2014215, DOE HEP QuantISED award \#100495, and the Gordon and Betty Moore Foundation Grant GBMF7946.
\end{acknowledgments}




\bibliography{Flavour_FOPT_biblio.bib}

\end{document}